\documentclass[manuscript,screen]{acmart}

\AtBeginDocument{%
 \providecommand\BibTeX{{%
 \normalfont B\kern-0.5em{\scshape i\kern-0.25em b}\kern-0.8em\TeX}}}

\copyrightyear{2024}
\acmConference[arXiv]{arXiv}{2024}{}

\begin{document}

\title{Tappy Plugin for Figma: Predicting Tap Success Rates of User-Interface Elements under Development for Smartphones}

\author{Shota Yamanaka}\email{syamanak@lycorp.co.jp}
\orcid{0000-0001-9807-120X}
\affiliation{\institution{LY Corporation}\city{Tokyo}\country{Japan}}

\author{Hiroki Usuba}\email{c-hiusuba@lycorp.jp}
\affiliation{\institution{LY Corporation}\city{Tokyo}\country{Japan}}

\author{Junichi Sato}\email{jsato@lycorp.jp}
\affiliation{\institution{LY Corporation}\city{Tokyo}\country{Japan}}
\authornote{He presently belongs to SB Intuitions Corp.}

\author{Naomi Sasaya}\email{nasasaya@lycorp.jp}
\affiliation{\institution{LY Corporation}\city{Tokyo}\country{Japan}}

\author{Fumiya Yamashita}\email{fyamashi@lycorp.co.jp}
\affiliation{\institution{LY Corporation}\city{Tokyo}\country{Japan}}

\author{Shuji Yamaguchi}\email{shyamagu@lycorp.co.jp}
\affiliation{\institution{LY Corporation}\city{Tokyo}\country{Japan}}

\renewcommand{\shortauthors}{Yamanaka et al.}
\renewcommand{\shorttitle}{Tappy Plugin for Figma}

\begin{abstract}
Tapping buttons and hyperlinks on smartphones is a fundamental operation, but users sometimes fail to tap user-interface (UI) elements.
Such mistakes degrade usability, and thus it is important for designers to configure UI elements so that users can accurately select them.
To support designers in setting a UI element with an intended tap success rate, we developed a plugin for Figma, which is modern software for developing webpages and applications for smartphones, based on our previously launched web-based application, \textit{Tappy}.
This plugin converts the size of a UI element from pixels to mm and then computes the tap success rates based on the Dual Gaussian Distribution Model.
We have made this plugin freely available to external users, so readers can install the Tappy plugin for Figma by visiting its installation page (\url{https://www.figma.com/community/plugin/1425006564066437139/tappy}) or from their desktop Figma software.
\end{abstract}

\ccsdesc[300]{Human-centered computing~Interactive systems and tools}

\keywords{Human motor performance, error rate prediction, endpoint distribution}

\maketitle

\section{Introduction}
User interface (UI) designers recognize that smartphone users sometimes mistakenly tap outside the intended elements, such as buttons, hyperlinks, link-embedded pictures, etc.
Such errors have a significant negative effect on usability~\cite{Banovic13,Yamanaka18mobilehci}.
Textbooks on UI design have thus suggested making UI elements large enough to tap easily~\cite{Clark16book,Hoober11book,Johnson14book,Neil14book}.

To predict how accurately users tap UI elements on webpages, we have launched a web-based application called Tappy (\url{https://tappy.yahoo.co.jp}) \cite{usuba24arxiv}. This application computes the tap success rate in \% for each UI element on a page. Tappy is designed for understanding tap accuracy for existing webpages, but we assume that predicting tap success rates is also useful when designers need to determine UI element sizes for their applications and webpages that are in development.

We therefore implemented a plugin for Figma\footnote{\url{https://www.figma.com/}} to provide the tap success rate prediction feature of Tappy.
Figma is widely used, modern software for designing applications and webpages for smartphones, PCs, and tablets.
We developed and launched this plugin for internal use within our corporation and publicly released it for external users on October 9th, 2024.
By October 30th, 2024, a total of 79 users had installed this plugin.

Researchers in the field of human-computer interaction (HCI) have derived numerous models related to UI operations (e.g., variations of Fitts' law \cite{Accot03,Bi13a}, success-rate models \cite{Bi16,Usuba22iss}).
Although it is typically claimed that such models are useful for designing UIs \cite{Yamanaka20issFFF,Zhang23shape,Huang20cross}, developing and launching a tool that utilizes these models to support professional designers' work is rare.
Because the Tappy plugin for Figma has been used within our corporation, we have evidence that professionals have benefited from this plugin.
In this paper, we explain the theoretical background of the Tappy plugin and how it works in Figma.


\section{Related Work}
\label{sec:RW}

When designers determine UI element sizes for touchscreens, there are guidelines to follow.
For example, the Android design guidelines recommend UI elements to be 9~mm or larger for usability~\cite{Android23}.
However, size restrictions may occur when designers need to arrange many elements on a single screen.
Quantitative models to predict tap success rates are useful for computing the decrease in accuracy in such cases~\cite{Bi16,Yamanaka20issFFF}.

Several tools that take into account implementation restrictions have been proposed to support and automate the design of applications and webpages.
Jiang et al.'s survey is quite detailed, and we recommend that readers refer to it~\cite{Jiang19orc}.
However, very few models for predicting usability metrics have been made available as design support tools. One example is our Tappy, which is an application that allows designers to analyze the tap success rates of UI elements in existing webpages~\cite{usuba24arxiv}.
Yet, it cannot calculate the tap success rates of the UI elements of applications or webpages under design.

Our web-based Tappy computes tap success rates from the sizes of UI elements.
According to Bi et al.'s \textit{Dual Gaussian Distribution Model}, the tap-point variance $\sigma^2$ and the square of the target size are linearly related for each of the x- and y-axes independently~\cite{Bi13a,Bi16}.
Assuming that the tap coordinates are random variables that follow normal distributions~\cite{Bi13a,Yamanaka23HPO}, the Tappy plugin we present in this paper uses Yamanaka and Usuba's model coefficients to predict a tap success rate for a rectangular target~\cite{Yamanaka24iss}:
\begin{equation}
 \label{eqs:srDefined}
 \mathrm{Success\ rate} = \text{erf}\left( \frac{W}{2 \sqrt{2} \sigma_x} \right) \text{erf}\left( \frac{H}{2 \sqrt{2} \sigma_y} \right),\ \ \mathrm{where}\ \ \ \sigma_x = \sqrt{0.0149 W^2 + 0.9414} \ \ \text{and}\ \ 
 \sigma_y = \sqrt{0.0091 H^2 + 1.0949},
\end{equation}
where $W$ and $H$ are the target sizes on the x- and y-axes in mm, respectively.


\section{Tappy Plugin for Figma}
\begin{figure}[t]
 \centering
 \includegraphics[width=0.8\textwidth]{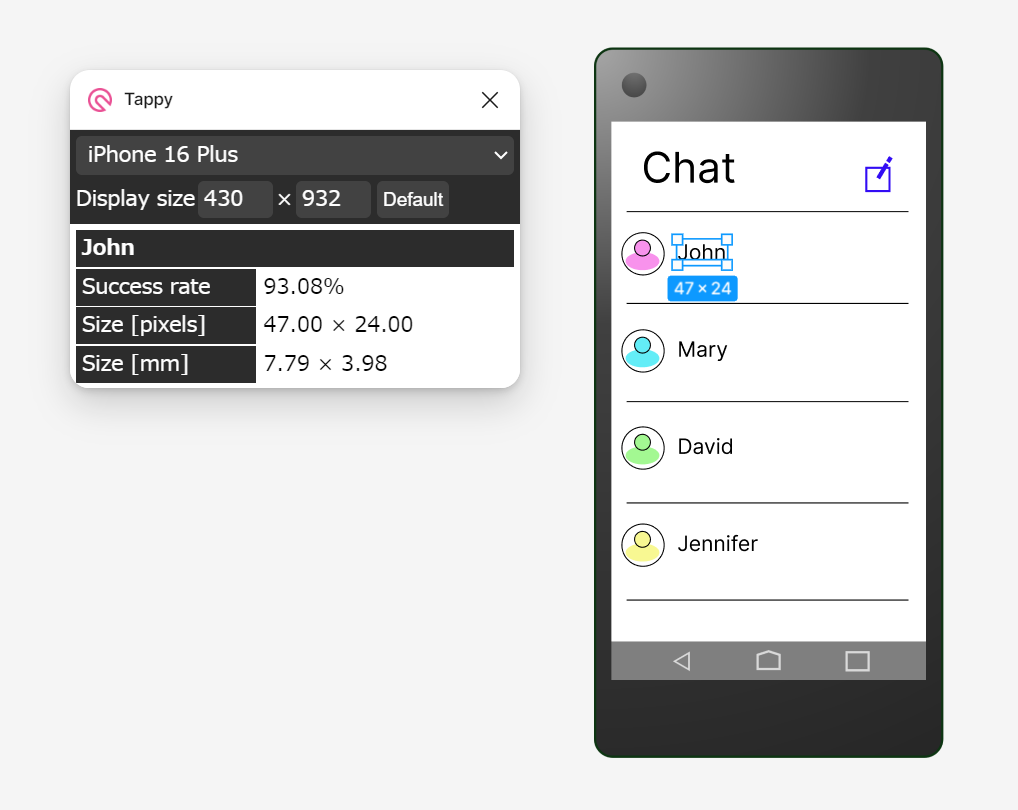}
 \caption{Screenshot of the Tappy plugin.}
 \label{fig:TappyFigma}
\end{figure}

To install the Tappy plugin, users can access the plugin introduction page (\url{https://www.figma.com/community/plugin/1425006564066437139/tappy}) and click the ``Open in...'' button, or search for \textit{Tappy} in ``Actions $\rightarrow$ Plugins \& widgets'' on Figma's web application or desktop software.
When users launch the Tappy plugin, a small floating window appears in Figma (Figure~\ref{fig:TappyFigma}).
This plugin currently supports iOS devices up to the iPhone 16 series.
In the Tappy plugin window, users select the target iOS device from a drop-down menu and click on a UI element in the application or webpage under development.
Then, the size of that element (in pixels and mm) and the tap success rate (in \%) are displayed in the window.

The Tappy plugin is implemented in JavaScript, and the size of the selected UI element is provided in pixels within the plugin implementation environment provided by Figma.
Because the screen resolutions (pixels per inch) and scale factors of all iOS devices are known, the Tappy plugin computes how large the selected UI element will actually be displayed on the screen in mm.
Then, using Equation~\ref{eqs:srDefined}, the tap success rate is predicted based on the physical size of that element.

\section{Future Work and Conclusion}
We developed a plugin for Figma to provide the Tappy feature, i.e., predicting tap success rates of UI elements for smartphone applications and webpages.
To predict success rates, we use data from a crowdsourced experiment where participants held the smartphone with their non-dominant hand and tapped targets with the index finger of their dominant hand~\cite{Yamanaka24iss}.
In addition, the actual success rate may vary depending on the position of the intended UI element on the screen~\cite{UsubaISS2023} and the arrangement of surrounding elements~\cite{Yamanaka18mobilehci,YamanakaISS2019}.
Updating the model to account for these effects is included in our future work.

In the Tappy plugin, the shape of a UI element is approximated as a bounding rectangle to simplify the computation of tap success rates.
However, element shapes can be arbitrary in actual smartphone applications and webpages.
The models for predicting the success rates of tapping circular elements~\cite{Yamanaka20issFFF,Bi16} and arbitrarily shaped elements~\cite{Zhang20moving} will be useful for future improvements to the Tappy plugin.
We hope that the human-performance models developed in the HCI field will be provided as practical tools so that many users can benefit from them.

\bibliographystyle{ACM-Reference-Format}
\bibliography{sample-base}


\begin{thebibliography}{21}


\ifx \showCODEN    \undefined \def \showCODEN     #1{\unskip}     \fi
\ifx \showDOI      \undefined \def \showDOI       #1{#1}\fi
\ifx \showISBNx    \undefined \def \showISBNx     #1{\unskip}     \fi
\ifx \showISBNxiii \undefined \def \showISBNxiii  #1{\unskip}     \fi
\ifx \showISSN     \undefined \def \showISSN      #1{\unskip}     \fi
\ifx \showLCCN     \undefined \def \showLCCN      #1{\unskip}     \fi
\ifx \shownote     \undefined \def \shownote      #1{#1}          \fi
\ifx \showarticletitle \undefined \def \showarticletitle #1{#1}   \fi
\ifx \showURL      \undefined \def \showURL       {\relax}        \fi
\providecommand\bibfield[2]{#2}
\providecommand\bibinfo[2]{#2}
\providecommand\natexlab[1]{#1}
\providecommand\showeprint[2][]{arXiv:#2}

\bibitem[Accot and Zhai(2003)]%
        {Accot03}
\bibfield{author}{\bibinfo{person}{Johnny Accot} {and} \bibinfo{person}{Shumin Zhai}.} \bibinfo{year}{2003}\natexlab{}.
\newblock \showarticletitle{Refining Fitts' Law Models for Bivariate Pointing}. In \bibinfo{booktitle}{\emph{Proceedings of the SIGCHI Conference on Human Factors in Computing Systems}} (Ft. Lauderdale, Florida, USA) \emph{(\bibinfo{series}{CHI '03})}. \bibinfo{publisher}{ACM}, \bibinfo{address}{New York, NY, USA}, \bibinfo{pages}{193--200}.
\newblock
\showISBNx{1-58113-630-7}
\urldef\tempurl%
\url{https://doi.org/10.1145/642611.642646}
\showDOI{\tempurl}


\bibitem[Banovic et~al\mbox{.}(2013)]%
        {Banovic13}
\bibfield{author}{\bibinfo{person}{Nikola Banovic}, \bibinfo{person}{Tovi Grossman}, {and} \bibinfo{person}{George Fitzmaurice}.} \bibinfo{year}{2013}\natexlab{}.
\newblock \showarticletitle{The Effect of Time-based Cost of Error in Target-directed Pointing Tasks}. In \bibinfo{booktitle}{\emph{Proceedings of the SIGCHI Conference on Human Factors in Computing Systems}} (Paris, France) \emph{(\bibinfo{series}{CHI '13})}. \bibinfo{publisher}{ACM}, \bibinfo{address}{New York, NY, USA}, \bibinfo{pages}{1373--1382}.
\newblock
\showISBNx{978-1-4503-1899-0}
\urldef\tempurl%
\url{https://doi.org/10.1145/2470654.2466181}
\showDOI{\tempurl}


\bibitem[Bi et~al\mbox{.}(2013)]%
        {Bi13a}
\bibfield{author}{\bibinfo{person}{Xiaojun Bi}, \bibinfo{person}{Yang Li}, {and} \bibinfo{person}{Shumin Zhai}.} \bibinfo{year}{2013}\natexlab{}.
\newblock \showarticletitle{FFitts Law: Modeling Finger Touch with Fitts' Law}. In \bibinfo{booktitle}{\emph{Proceedings of the SIGCHI Conference on Human Factors in Computing Systems}} (Paris, France) \emph{(\bibinfo{series}{CHI '13})}. \bibinfo{publisher}{ACM}, \bibinfo{address}{New York, NY, USA}, \bibinfo{pages}{1363--1372}.
\newblock
\showISBNx{978-1-4503-1899-0}
\urldef\tempurl%
\url{https://doi.org/10.1145/2470654.2466180}
\showDOI{\tempurl}


\bibitem[Bi and Zhai(2016)]%
        {Bi16}
\bibfield{author}{\bibinfo{person}{Xiaojun Bi} {and} \bibinfo{person}{Shumin Zhai}.} \bibinfo{year}{2016}\natexlab{}.
\newblock \showarticletitle{Predicting Finger-Touch Accuracy Based on the Dual Gaussian Distribution Model}. In \bibinfo{booktitle}{\emph{Proceedings of the 29th Annual Symposium on User Interface Software and Technology}} (Tokyo, Japan) \emph{(\bibinfo{series}{UIST '16})}. \bibinfo{publisher}{ACM}, \bibinfo{address}{New York, NY, USA}, \bibinfo{pages}{313--319}.
\newblock
\showISBNx{978-1-4503-4189-9}
\urldef\tempurl%
\url{https://doi.org/10.1145/2984511.2984546}
\showDOI{\tempurl}


\bibitem[Clark(2016)]%
        {Clark16book}
\bibfield{author}{\bibinfo{person}{Josh Clark}.} \bibinfo{year}{2016}\natexlab{}.
\newblock \bibinfo{booktitle}{\emph{Designing for Touch} (\bibinfo{edition}{1} ed.)}.
\newblock \bibinfo{publisher}{A Book Apart}, \bibinfo{address}{New York, USA}.
\newblock
\showISBNx{9781937557294}


\bibitem[Google(2023)]%
        {Android23}
\bibfield{author}{\bibinfo{person}{Google}.} \bibinfo{year}{2023}\natexlab{}.
\newblock \bibinfo{title}{Touch target size}.
\newblock
\newblock
\urldef\tempurl%
\url{https://support.google.com/accessibility/android/answer/7101858?hl=en}
\showURL{%
\tempurl}


\bibitem[Hoober and Berkman(2011)]%
        {Hoober11book}
\bibfield{author}{\bibinfo{person}{Steven Hoober} {and} \bibinfo{person}{Eric Berkman}.} \bibinfo{year}{2011}\natexlab{}.
\newblock \bibinfo{booktitle}{\emph{Designing Mobile Interfaces: Patterns for Interaction Design} (\bibinfo{edition}{1} ed.)}.
\newblock \bibinfo{publisher}{O'Reilly Media, Inc.}, \bibinfo{address}{1005 Gravenstein Highway North, Sebastopol, CA 95472, USA}.
\newblock
\showISBNx{9781449394639}


\bibitem[Huang et~al\mbox{.}(2020)]%
        {Huang20cross}
\bibfield{author}{\bibinfo{person}{Jin Huang}, \bibinfo{person}{Feng Tian}, \bibinfo{person}{Xiangmin Fan}, \bibinfo{person}{Huawei Tu}, \bibinfo{person}{Hao Zhang}, \bibinfo{person}{Xiaolan Peng}, {and} \bibinfo{person}{Hongan Wang}.} \bibinfo{year}{2020}\natexlab{}.
\newblock \bibinfo{booktitle}{\emph{Modeling the Endpoint Uncertainty in Crossing-Based Moving Target Selection}}.
\newblock \bibinfo{publisher}{Association for Computing Machinery}, \bibinfo{address}{New York, NY, USA}, \bibinfo{pages}{1–12}.
\newblock
\showISBNx{9781450367080}
\urldef\tempurl%
\url{https://doi.org/10.1145/3313831.3376336}
\showURL{%
\tempurl}


\bibitem[Jiang et~al\mbox{.}(2019)]%
        {Jiang19orc}
\bibfield{author}{\bibinfo{person}{Yue Jiang}, \bibinfo{person}{Ruofei Du}, \bibinfo{person}{Christof Lutteroth}, {and} \bibinfo{person}{Wolfgang Stuerzlinger}.} \bibinfo{year}{2019}\natexlab{}.
\newblock \showarticletitle{ORC Layout: Adaptive GUI Layout with OR-Constraints}. In \bibinfo{booktitle}{\emph{Proceedings of the 2019 CHI Conference on Human Factors in Computing Systems}} (Glasgow, Scotland Uk) \emph{(\bibinfo{series}{CHI '19})}. \bibinfo{publisher}{Association for Computing Machinery}, \bibinfo{address}{New York, NY, USA}, \bibinfo{pages}{1–12}.
\newblock
\showISBNx{9781450359702}
\urldef\tempurl%
\url{https://doi.org/10.1145/3290605.3300643}
\showDOI{\tempurl}


\bibitem[Johnson(2014)]%
        {Johnson14book}
\bibfield{author}{\bibinfo{person}{Jeff Johnson}.} \bibinfo{year}{2014}\natexlab{}.
\newblock \bibinfo{booktitle}{\emph{Designing with the Mind in Mind: Simple Guide to Understanding User Interface Design Guidelines} (\bibinfo{edition}{2} ed.)}.
\newblock \bibinfo{publisher}{Morgan Kaufmann}, \bibinfo{address}{San Francisco, USA}.
\newblock
\showISBNx{978-0-12-407914-4}
\urldef\tempurl%
\url{https://doi.org/10.1016/C2012-0-07128-1}
\showDOI{\tempurl}


\bibitem[Neil(2014)]%
        {Neil14book}
\bibfield{author}{\bibinfo{person}{Theresa Neil}.} \bibinfo{year}{2014}\natexlab{}.
\newblock \bibinfo{booktitle}{\emph{Mobile Design Pattern Gallery: UI Patterns for Smartphone Apps} (\bibinfo{edition}{1} ed.)}.
\newblock \bibinfo{publisher}{O'Reilly Media, Inc.}, \bibinfo{address}{1005 Gravenstein Highway North, Sebastopol, CA 95472, USA}.
\newblock
\showISBNx{978-1-4493-6363-5}


\bibitem[Usuba et~al\mbox{.}(2024)]%
        {usuba24arxiv}
\bibfield{author}{\bibinfo{person}{Hiroki Usuba}, \bibinfo{person}{Junichi Sato}, \bibinfo{person}{Naomi Sasaya}, \bibinfo{person}{Shota Yamanaka}, {and} \bibinfo{person}{Fumiya Yamashita}.} \bibinfo{year}{2024}\natexlab{}.
\newblock \bibinfo{title}{Tappy: Predicting Tap Accuracy of User-Interface Elements by Reverse-Engineering Webpage Structures}.
\newblock
\newblock
\showeprint[arxiv]{2403.03097}~[cs.HC]
\urldef\tempurl%
\url{https://arxiv.org/abs/2403.03097}
\showURL{%
\tempurl}


\bibitem[Usuba et~al\mbox{.}(2023)]%
        {UsubaISS2023}
\bibfield{author}{\bibinfo{person}{Hiroki Usuba}, \bibinfo{person}{Shota Yamanaka}, {and} \bibinfo{person}{Junichi Sato}.} \bibinfo{year}{2023}\natexlab{}.
\newblock \showarticletitle{Clarifying the Effect of Edge Targets in Touch Pointing through Crowdsourced Experiments}.
\newblock \bibinfo{journal}{\emph{Proc. ACM Hum.-Comput. Interact.}} \bibinfo{volume}{7}, \bibinfo{number}{ISS}, Article \bibinfo{articleno}{433} (\bibinfo{date}{nov} \bibinfo{year}{2023}), \bibinfo{numpages}{19}~pages.
\newblock
\urldef\tempurl%
\url{https://doi.org/10.1145/3626469}
\showDOI{\tempurl}


\bibitem[Usuba et~al\mbox{.}(2022)]%
        {Usuba22iss}
\bibfield{author}{\bibinfo{person}{Hiroki Usuba}, \bibinfo{person}{Shota Yamanaka}, \bibinfo{person}{Junichi Sato}, {and} \bibinfo{person}{Homei Miyashita}.} \bibinfo{year}{2022}\natexlab{}.
\newblock \showarticletitle{Predicting Touch Accuracy for Rectangular Targets by Using One-Dimensional Task Results}.
\newblock \bibinfo{journal}{\emph{Proc. ACM Hum.-Comput. Interact.}} \bibinfo{volume}{6}, \bibinfo{number}{ISS}, Article \bibinfo{articleno}{579} (\bibinfo{date}{nov} \bibinfo{year}{2022}), \bibinfo{numpages}{13}~pages.
\newblock
\urldef\tempurl%
\url{https://doi.org/10.1145/3567732}
\showDOI{\tempurl}


\bibitem[Yamanaka(2018)]%
        {Yamanaka18mobilehci}
\bibfield{author}{\bibinfo{person}{Shota Yamanaka}.} \bibinfo{year}{2018}\natexlab{}.
\newblock \showarticletitle{Effect of Gaps with Penal Distractors Imposing Time Penalty in Touch-pointing Tasks}. In \bibinfo{booktitle}{\emph{Proceedings of the 20th International Conference on Human-Computer Interaction with Mobile Devices and Services}} (Barcelona, Spain) \emph{(\bibinfo{series}{MobileHCI '18})}. \bibinfo{publisher}{ACM}, \bibinfo{address}{New York, NY, USA}, \bibinfo{numpages}{8}~pages.
\newblock
\showISBNx{978-1-4503-5898-9/18/09}
\urldef\tempurl%
\url{https://doi.org/10.1145/3229434.3229435}
\showDOI{\tempurl}


\bibitem[Yamanaka et~al\mbox{.}(2019)]%
        {YamanakaISS2019}
\bibfield{author}{\bibinfo{person}{Shota Yamanaka}, \bibinfo{person}{Hiroaki Shimono}, {and} \bibinfo{person}{Homei Miyashita}.} \bibinfo{year}{2019}\natexlab{}.
\newblock \showarticletitle{Towards More Practical Spacing for Smartphone Touch GUI Objects Accompanied by Distractors}. In \bibinfo{booktitle}{\emph{Proceedings of the 2019 ACM International Conference on Interactive Surfaces and Spaces}} \emph{(\bibinfo{series}{ISS '19})}. \bibinfo{publisher}{Association for Computing Machinery}, \bibinfo{address}{New York, NY, USA}, \bibinfo{pages}{157–169}.
\newblock
\showISBNx{9781450368919}
\urldef\tempurl%
\url{https://doi.org/10.1145/3343055.3359698}
\showDOI{\tempurl}


\bibitem[Yamanaka and Usuba(2020)]%
        {Yamanaka20issFFF}
\bibfield{author}{\bibinfo{person}{Shota Yamanaka} {and} \bibinfo{person}{Hiroki Usuba}.} \bibinfo{year}{2020}\natexlab{}.
\newblock \showarticletitle{Rethinking the Dual Gaussian Distribution Model for Predicting Touch Accuracy in On-Screen-Start Pointing Tasks}.
\newblock \bibinfo{journal}{\emph{Proc. ACM Hum.-Comput. Interact.}} \bibinfo{volume}{4}, \bibinfo{number}{ISS}, Article \bibinfo{articleno}{205} (\bibinfo{date}{Nov.} \bibinfo{year}{2020}), \bibinfo{numpages}{20}~pages.
\newblock
\urldef\tempurl%
\url{https://doi.org/10.1145/3427333}
\showDOI{\tempurl}


\bibitem[Yamanaka and Usuba(2023)]%
        {Yamanaka23HPO}
\bibfield{author}{\bibinfo{person}{Shota Yamanaka} {and} \bibinfo{person}{Hiroki Usuba}.} \bibinfo{year}{2023}\natexlab{}.
\newblock \showarticletitle{Tuning Endpoint-variability Parameters by Observed Error Rates to Obtain Better Prediction Accuracy of Pointing Misses}. In \bibinfo{booktitle}{\emph{Proceedings of the 2023 CHI Conference on Human Factors in Computing Systems}} \emph{(\bibinfo{series}{CHI '23})}. \bibinfo{publisher}{Association for Computing Machinery}, \bibinfo{address}{New York, NY, USA}, Article \bibinfo{articleno}{579}, \bibinfo{numpages}{18}~pages.
\newblock
\showISBNx{9781450394215}
\urldef\tempurl%
\url{https://doi.org/10.1145/3544548.3580746}
\showDOI{\tempurl}


\bibitem[Yamanaka and Usuba(2024)]%
        {Yamanaka24iss}
\bibfield{author}{\bibinfo{person}{Shota Yamanaka} {and} \bibinfo{person}{Hiroki Usuba}.} \bibinfo{year}{2024}\natexlab{}.
\newblock \showarticletitle{0.2-mm-Step Verification of the Dual Gaussian Distribution Model with Large Sample Size for Predicting Tap Success Rates}.
\newblock \bibinfo{journal}{\emph{Proc. ACM Hum.-Comput. Interact.}} \bibinfo{volume}{8}, \bibinfo{number}{ISS}, Article \bibinfo{articleno}{553} (\bibinfo{date}{Oct.} \bibinfo{year}{2024}), \bibinfo{numpages}{20}~pages.
\newblock
\urldef\tempurl%
\url{https://doi.org/10.1145/3698153}
\showDOI{\tempurl}


\bibitem[Zhang et~al\mbox{.}(2023)]%
        {Zhang23shape}
\bibfield{author}{\bibinfo{person}{Hao Zhang}, \bibinfo{person}{Jin Huang}, \bibinfo{person}{Huawei Tu}, {and} \bibinfo{person}{Feng Tian}.} \bibinfo{year}{2023}\natexlab{}.
\newblock \showarticletitle{Shape-Adaptive Ternary-Gaussian Model: Modeling Pointing Uncertainty for Moving Targets of Arbitrary Shapes}. In \bibinfo{booktitle}{\emph{Proceedings of the 2023 CHI Conference on Human Factors in Computing Systems}} \emph{(\bibinfo{series}{CHI '23})}. \bibinfo{publisher}{Association for Computing Machinery}, \bibinfo{address}{New York, NY, USA}, Article \bibinfo{articleno}{777}, \bibinfo{numpages}{18}~pages.
\newblock
\showISBNx{9781450394215}
\urldef\tempurl%
\url{https://doi.org/10.1145/3544548.3581217}
\showDOI{\tempurl}


\bibitem[Zhang et~al\mbox{.}(2020)]%
        {Zhang20moving}
\bibfield{author}{\bibinfo{person}{Ziyue Zhang}, \bibinfo{person}{Jin Huang}, {and} \bibinfo{person}{Feng Tian}.} \bibinfo{year}{2020}\natexlab{}.
\newblock \showarticletitle{Modeling the Uncertainty in Pointing of Moving Targets with Arbitrary Shapes}. In \bibinfo{booktitle}{\emph{Extended Abstracts of the 2020 CHI Conference on Human Factors in Computing Systems}} (Honolulu, HI, USA) \emph{(\bibinfo{series}{CHI EA '20})}. \bibinfo{publisher}{Association for Computing Machinery}, \bibinfo{address}{New York, NY, USA}, \bibinfo{pages}{1–7}.
\newblock
\showISBNx{9781450368193}
\urldef\tempurl%
\url{https://doi.org/10.1145/3334480.3382875}
\showDOI{\tempurl}


\end{thebibliography}

\end{document}